\documentclass[a4paper,fleqn,usenatbib]{mnras}

\usepackage[T1]{fontenc}
\usepackage{ae,aecompl}

\usepackage{graphicx}   
\usepackage{amsmath}    
\usepackage{amssymb}    

\newcommand{\be}{\begin{equation}}

\newcommand{\ee}{\end{equation}}
\newcommand{\bea}{\begin{eqnarray}}
\newcommand{\eea}{\end{eqnarray}}
\newcommand{\bef}{\begin{figure}}
\newcommand{\eef}{\end{figure}}
\newcommand{\bce}{\begin{center}}
\newcommand{\ece}{\end{center}}

\def\lsim{\mathrel{\rlap{\lower4pt\hbox{\hskip1pt$\sim$}}
    \raise1pt\hbox{$<$}}}         
\def\gsim{\mathrel{\rlap{\lower4pt\hbox{\hskip1pt$\sim$}}
    \raise1pt\hbox{$>$}}}         
\title[Ellipticity of a neutron star]{The permanent ellipticity of the neutron star in PSR J1023+0038}

\author[Bhattacharyya]{Sudip Bhattacharyya\thanks{E-mail: sudip@tifr.res.in}\\
       Department of Astronomy and Astrophysics, Tata Institute of Fundamental Research, Mumbai 400005, India}

\date{Accepted 2020 July 30. Received 2020 July 22; in original form 2020 June 11}

\pubyear{2015}

\begin{document}
\label{firstpage}
\pagerange{\pageref{firstpage}--\pageref{lastpage}}
\maketitle

\begin{abstract}
A millisecond pulsar having an ellipticity, that is an asymmetric mass 
distribution around its spin-axis, could emit continuous gravitational waves,
which have not been detected so far.
An indirect way to infer such waves is to estimate the contribution of
the waves to the spin-down rate of the pulsar. The transitional pulsar 
PSR J1023+0038 is ideal and unique for this purpose, because this is the only
millisecond pulsar for which the spin-down rate has been measured in
both accreting and non-accreting states. Here we infer, from our formalism
based on the complete torque 
budget equations and the pulsar magnetospheric origin of observed
$\gamma$-rays in the two states, that PSR J1023+0038 should emit gravitational
waves due to a permanent ellipticity of the pulsar. 
The formalism also explains some other main observational aspects of
this source in a self-consistent way.
As an example, our formalism naturally infers the accretion disc penetration into 
the pulsar magnetosphere, and explains the observed X-ray pulsations in the 
accreting state using the standard and well-accepted scenario. 
This, in turn, infers the larger pulsar spin-down power in the accreting state, which,
in our formalism, explains the observed larger $\gamma$-ray emission in this state.
Exploring wide ranges of parameter values of PSR J1023+0038, and not assuming an 
additional source of stellar ellipticity in the accreting state, we find
the misaligned mass quadrupole moment of the pulsar in the range of
$(0.92-1.88)\times10^{36}$~g cm$^2$, implying an ellipticity range of
$(0.48-0.93)\times10^{-9}$.
\end{abstract}

\begin{keywords}
accretion, accretion discs --- gravitational waves --- pulsars: general --- pulsars: individual (PSR J1023+0038) --- stars: neutron --- X-rays: binaries
\end{keywords}



\section{Introduction}\label{Introduction}

Observations and theoretical studies have indicated that millisecond pulsars (MSPs), i.e.,
rapidly spinning neutron stars which show periodic brightness oscillations, can
have a misaligned mass distribution or ellipticity ($\epsilon \sim 10^{-9}$) around the spin-axis,
and hence could emit gravitational waves continuously
\citep{Bildsten1998,Anderssonetal1999,Chakrabartyetal2003,BhattacharyyaChakrabarty2017,Bhattacharyya2017,Woanetal2018}.
But, so far only upper limits ($> 10^{-8}$) of ellipticity could be estimated using
the Laser Interferometer Gravitational-Wave Observatory \citep[LIGO; ][]{Abbottetal2019}.
However, a pulsar's spin frequency ($\nu$) and its rate of change ($\dot \nu$)
can be measured from electromagnetic (e.g., radio, X-ray) observations,
and an alternative way to infer gravitational waves is to
estimate the contribution of these waves to the $\dot \nu$.
While such an estimation could not be reliably achieved so far,
PSR J1023+0038 (hereafter, J1023) is an ideal pulsar for this study using torque
($= 2\pi \dot \nu I$; $I$ is the stellar moment of inertia) budget calculations.
This is because J1023 is one of the three known transitional MSPs,
which are observed to swing between the non-accreting spin-powered radio MSP
(RMSP) state and the accreting low-mass X-ray binary (LMXB) state 
\citep{Archibaldetal2009}. Moreover, J1023 is
unique because its $\dot \nu$ has been measured in both RMSP and LMXB states.
In fact, based on the difference in $\dot \nu$ values in
these two states, a model involving a stellar ellipticity only in
the LMXB state was proposed \citep{HaskellPatruno2017}.

But a torque budget has hitherto not been systematically studied for J1023, 
although several classes of models have been proposed to explain its observational 
aspects. Here we briefly mention some of the noteworthy observational features
and estimated parameter values of J1023.
The spin frequency, mass ($M$) and distance of this pulsar or the neutron star,
which may have an unevolved and Roche-lobe filling low-mass companion star of mass $\sim 0.2 M_\odot$
\citep{ThorstensenArmstrong2005,Archibaldetal2010,Tendulkaretal2014}, are $592.42$~Hz,
$1.71\pm0.16 M_\odot$ and $1368_{-39}^{+42}$~pc respectively \citep{Archibaldetal2009,Delleretal2012}.
In the RMSP state, the companion star was found to be irradiated by the neutron star, which gave rise
to an optical light curve with the orbital period of $\sim 0.198094$ days and with a peak-to-peak
amplitude of $\sim 0.3$~magnitude, or a half-amplitude of $\sim 180$~K around a $\sim$~G6 type star
\citep{ThorstensenArmstrong2005}. Using a total binary system mass of $1.91 M_\odot$ and
the measured orbital period, the binary separation $a$ can be estimated.
A $180$~K half-amplitude around the $5590$~K temperature of a G6 type companion star
gives maximum and minimum temperatures: $T_{\rm max} = 5770$~K and $T_{\rm min} = 5410$~K.
Then using the formula $4\pi a^2 \sigma (T_{\rm max}^4 - T_{\rm min}^4)$
($\sigma$ is the Stefan-Boltzmann constant),
one can estimate a pulsar isotropic luminosity of at least $2.7\times10^{33}$ erg s$^{-1}$.
The irradiation was found to be similar in the LMXB state \citep{Kennedyetal2018}.

The measured spin-down rate $\dot{\nu}$ of J1023 in the RMSP state is
$\approx -2.40\times10^{-15}$~Hz s$^{-1}$ 
\citep{Delleretal2012,Archibaldetal2013,Jaodandetal2016}. 
However, considering the Shklovskii
effect and the net effect of acceleration in the Galactic gravitational potential,
the intrinsic spin-down rate in the RMSP state was estimated to be
$\dot{\nu}_{\rm RMSP} \approx -1.89\times10^{-15}$~Hz s$^{-1}$,
which for a moment of inertia of $10^{45}$ g cm$^2$ gives a spin-down 
luminosity of $\sim 4.4\times10^{34}$ erg s$^{-1}$ 
\citep{Delleretal2012,Archibaldetal2013,PapittoTorres2015}.
In the RMSP state, a $\gamma$-ray luminosity ($L_{\gamma, \rm{RMSP}}$) of
$\sim 1.2\times10^{33}$ erg s$^{-1}$ was detected \citep{Nolanetal2012,PapittoTorres2015}.
In the LMXB state, the $\gamma$-ray flux was observed to increase by a factor of 
$\sim 5-6.5$ relative to that in the RMSP state 
\citep{Stappersetal2014, Delleretal2015}.
While this might imply a luminosity $L_{\gamma, \rm{LMXB}} \approx 
(6.0-7.8)\times10^{33}$ erg s$^{-1}$, \cite{Torresetal2017} reported 
$L_{\gamma, \rm{LMXB}} \approx 12.4\times10^{33}$ erg s$^{-1}$.
No radio pulsation was detected in the LMXB state \citep{Delleretal2015}.
The presence of an accretion disc in the LMXB state was confirmed by the broad double-peaked
H$\alpha$ emission line observed in the optical spectrum \citep{Wangetal2009}.
The LMXB state X-ray luminosity ($L_{\rm X}$), which is typically much higher than
that in the RMSP state, has three distinctly different modes:
low-mode ($L_{\rm X} \sim 10^{33}$ erg s$^{-1}$), high-mode
($L_{\rm X} \sim 7\times10^{33}$ erg s$^{-1}$) and flaring-mode
\citep[$L_{\rm X} \sim 2\times10^{34}$ erg s$^{-1}$; ][]{PapittoTorres2015,Campanaetal2016}.
The source typically spends $\sim 22$\%, $\sim 77$\% and $\sim 1$\% time in these modes respectively
\citep{Jaodandetal2016}, and can switch between modes in $\sim 10$~s \citep{Bogdanovetal2015}.
X-ray pulsations (rms amplitude $\sim 5-10$\%), i.e., a brightness variation of observed 
X-rays with the pulsar's spin period, have been detected only in the high-mode
\citep{Archibaldetal2015,Jaodandetal2016}.
The spin-down rate $\dot{\nu}$ of the pulsar in the LMXB state, measured from X-ray 
pulsations, is $\approx -3.04\times10^{-15}$~Hz s$^{-1}$ \citep{Jaodandetal2016}.
However, as for the RMSP state, the intrinsic spin-down rate in the LMXB state 
could be estimated to be $\dot{\nu}_{\rm LMXB}\approx 
-2.50\times10^{-15}$~Hz s$^{-1}$.
The X-ray behaviour of the source in low- and high-modes
is exceptionally stable \citep{Jaodandetal2016}. Apart from X-ray pulsations,
recently, optical pulsations at the pulsar spin frequency have also been discovered
from J1023 in high and flaring modes in the LMXB state \citep{Ambrosinoetal2017}.
This is the first and so far the only MSP to show such optical pulsations.
It was inferred that optical pulses were originated within a few kilometers
from the location of the X-ray pulse origin \citep{Papittoetal2019}.

In this paper, we systematically study the torque budget, the energy budget 
and some other observational aspects of J1023, in order to find if the 
neutron star has a non-zero ellipticity.
In section~\ref{Previous}, we discuss the main classes of models proposed to explain 
the observational aspects of J1023. Section~\ref{Formalism} describes our formalism
in a simple manner, as well as our primary results, and section~\ref{Feasibility}
discusses the feasibility of the 
formalism, and its implications. Together these two sections strongly suggest that the
neutron star in J1023 can have an ellipticity.
In section~\ref{Method}, we describe the method to constrain source
parameter values using our formalism, and in section~\ref{Constraints},
we report the constrained ranges of parameter values.
In section~\ref{Summary}, we summarize our results and conclusions.

\section{Previous models of PSR J1023+0038}\label{Previous}

In this section, we discuss some existing models of observational aspects 
of J1023. These not only could provide some motivations for our
work, but also can be useful to explain our formalism.

The X-ray pulsations of J1023 in the high-mode of the LMXB state are similar to such a
timing feature observed from accretion-powered millisecond X-ray pulsars (AMXPs). For AMXPs, X-ray pulsations are
well-understood, and here we briefly describe this standard and well-accepted scenario, which
requires an accretion disc. The disc is stopped
by the pulsar dipolar magnetosphere at the magnetospheric radius
\citep[][and references therein]{PatrunoWatts2012}
\begin{equation}\label{rm}
        r_{\rm m} = \xi \left(\frac{\mu^4}{2 G M \dot M^2}\right)^{1/7},
\end{equation}
where, $G$ is the gravitational constant, $\dot M$ is the accretion rate, $\mu~(= BR^3)$ is the pulsar
magnetic dipole moment, $B$ is the pulsar surface dipole magnetic field, $R$ is the
pulsar radius and $\xi$ is an order of unity constant, which depends on the threading
of the disc by the pulsar magnetic field lines \citep{Wang1996}.
In the `accretion regime', when $r_{\rm m}$ is less than the corotation radius \citep{PatrunoWatts2012}
\begin{equation}\label{rco}
        r_{\rm co} = \left(\frac{GM}{4\pi^2 \nu^2}\right)^{1/3},
\end{equation}
but greater than $R$ and $r_{\rm ISCO}$ (innermost stable circular orbit (ISCO) radius),
the accreted matter can be channelled to the pulsar's magnetic polar caps, giving rise to
X-ray pulsations.
On the other hand, for $r_{\rm lc} > r_{\rm m} > r_{\rm co}$, i.e., in the
`propeller regime', accreted matter can be partially or fully driven away from the system \citep{PatrunoWatts2012}.
Here, $r_{\rm lc} = c/2\pi\nu$ ($\approx 80.5$~km for J1023; $c$ is the speed of light in vacuum) is the light cylinder radius.

But, does this mechanism for AMXPs work for the X-ray pulsations observed from J1023?
Here, we mention independent observational indications. 
(1) J1023's X-ray pulsation characteristics are strikingly similar 
to those of the known AMXPs \citep{Archibaldetal2015}, which suggests
that they have the same pulsation mechanism.
(2) A disc component was $6.1\sigma$ significant in the high-mode X-ray spectrum of
J1023, as found from the {\it XMM-Newton} satellite data \citep{Campanaetal2016}. 
This spectral analysis gave a disc inner edge radius
($r_{\rm in}$) of $21_{-7}^{+9}$~km. From a later analysis with additional X-ray data, $r_{\rm in}$ was estimated
to be $26_{-9}^{+11}$~km \citep{CotiZelatietal2018}. Both the spectral results suggest that the accretion disc
could have extended inside the $r_{\rm co}$ ($\approx 24.6-26.2$~km, for
$M \approx 1.55-1.87 M_\odot$ for J1023) in the high mode, and hence the X-ray pulsations of J1023
might have the same origin as in other AMXPs.

However, if one assumes that the measured $\dot{\nu}_{\rm RMSP}$ is entirely caused by the pulsar magnetospheric activities,
the inferred high $\mu$ value ($\sim 10^{26}$~G~cm$^3$)
implies that $r_{\rm m}$ is greater than $r_{\rm co}$,
and hence the requirement for the standard X-ray pulsation mechanism is not satisfied.
An approach, which could allow a significantly lower value of $r_{\rm m}$, 
i.e., $r_{\rm m} < r_{\rm co}$
even for $\mu \sim 10^{26}$~G~cm$^3$ for J1023, was discussed in
\cite{Bozzoetal2018}. But this paper mentioned that such an approach could not explain other phenomena observed in the
LMXB state of J1023.
Therefore, almost none of the major existing models for J1023 can accommodate the
standard and well-accepted mechanism for X-ray pulsations in the LMXB state, as all of them
use a relatively high $\mu$ value inferred from the measured $\dot{\nu}_{\rm RMSP}$.
Now we briefly mention some classes of these existing models.

One class of such models assumes that the pulsar magnetospheric activities operate at the
same level in both RMSP and LMXB states. Therefore, this class of models cannot explain a significantly higher
$|\dot{\nu}_{\rm LMXB}|$ value compared to the $|\dot{\nu}_{\rm RMSP}|$ value, based
on such equal magnetospheric activities.
Some of these models attempt to explain observed X-ray properties
in the LMXB state in terms of the spin power of the pulsar \citep{Stappersetal2014,Takataetal2014}.
Howover, while these models
might explain the higher $\gamma$-ray luminosity value in the LMXB state,
most of them may not explain the
remarkably stable LMXB X-ray modes and fast switching between them, X-ray pulsations only in the high-mode, X-ray
pulse profiles \citep[see ][]{PapittoTorres2015}, and the measured 
low disc inner edge radius $r_{\rm in}$ \citep{Campanaetal2016}.

Another type of models involves a high accretion rate ($\dot M$),
about a hundred times the value inferred from X-ray observations,
in the high-mode of J1023 \citep{PapittoTorres2015}. According to such a model,
$r_{\rm in}$ is close to but greater than $r_{\rm co}$, and hence the source is in the propeller
regime in the high-mode \citep{Campanaetal2016}. Since matter is propelled away from the system
in the propeller regime, it is claimed that $\sim 99$\% of the disc matter is
accelerated and ejected from J1023 \citep{PapittoTorres2015}.
This might explain the $\sim 5-6.5$ times higher $\gamma$-ray luminosity in the
LMXB state. However, this would also imply a $|\dot{\nu}_{\rm LMXB}|$ value at 
least twice the $|\dot{\nu}_{\rm RMSP}|$ value \citep{PapittoTorres2015}, which 
is much higher than that observed \citep[see also][]{Papittoetal2019}. 
Note that this model might imply an even higher $|\dot{\nu}_{\rm LMXB}|$ value,
if $L_{\gamma, \rm{LMXB}} \approx 12.4\times10^{33}$ erg s$^{-1}$,
reported by \cite{Torresetal2017} (see section~\ref{Introduction}), is considered.
Besides, it may not be possible to expel 
$\sim 99$\% of the disc matter by propeller activities
when $r_{\rm in}$ is {\it not} much higher than $r_{\rm co}$ \citep[see ][]{Campanaetal2016},
as most of this matter would come back and accumulate, thus making the disc extend inside
$r_{\rm co}$ \citep{DAngeloetal2010}. This would result in a much higher $L_X$ than 
that observed, and a large spin-up torque which is not observed.
In this $\sim 99$\% accreted matter ejection model, even for $r_{\rm m} > r_{\rm co}$, 
the X-ray luminosity might be much higher than
the observed high-mode X-ray luminosity \citep[see ][]{Veledinaetal2019}.
Besides, such a model may not explain the remarkable stability of the high-mode,
as even a very small fractional change in the $\sim 99$\% expelled matter would 
change the $\sim 1$\% accreting matter by a large fraction.
Moreover, this model relies on the disc-magnetosphere interaction which should not drastically change
from one system to another, as long as the neutron star $\mu$ and $\nu$ values and the $\dot M$ value are
not very different. Therefore, if the accretion rate is indeed so high at such a low X-ray luminosity
of J1023, $\sim 99$\% of this accreted matter is propelled away to give rise to an enhanced
$\gamma$-ray luminosity,
and $\sim 1$\% of this accreted matter gives rise to X-ray pulsations,
we would expect to observe such features from the known AMXPs with similar $\mu$ values ($\sim 10^{26}$~G~cm$^3$)
at similarly low X-ray luminosities. But, these have not been observed.
Due to all these reasons, we conclude that an accretion rate much higher than the one indicated
by the observed X-ray luminosity, and the corresponding propeller effect, may not explain the
LMXB state of J1023.

Another type of model \citep{HaskellPatruno2017} attempts to explain the difference between $\dot{\nu}_{\rm RMSP}$ and
$\dot{\nu}_{\rm LMXB}$, assuming an extra spin-down torque due to gravitational waves emitted
from the neutron star {\it only} in the LMXB state. However, this paper \citep{HaskellPatruno2017} does not address other major observational
aspects of J1023 in the LMXB state, such as the higher $\gamma$-ray luminosity,
stability and fast switching
of modes, and X-ray pulsations only in the high-mode. Moreover, while this model requires accretion onto the
neutron star in the LMXB state, it also implicitly assumes a relatively high $\mu$ value inferred from
the measured $\dot{\nu}_{\rm RMSP}$. However, it is not discussed how the accreted matter would
reach the neutron star surface for such a high $\mu$ value.
Besides, if the accretion disc comes close to the neutron star, at least a fraction of
the magnetic field lines intersecting the disc should be opened
\citep[reported from simulations; see ][]{Parfreyetal2016, ParfreyetalTchekhovskoy2017},
and would likely increase the spin-down torque by several times, relative to the torque
in the RMSP state, due to enhanced pulsar magnetospheric activities
(section~\ref{gamma} discusses this point in more detail).
Note that here we consider these activities are not quenched in the LMXB state, as 
\cite{HaskellPatruno2017} implicitly assume.
Such an additional spin-down torque would imply a higher value of 
$|\dot{\nu}_{\rm LMXB}|$ than the one inferred (see section~\ref{Introduction}).
Moreover, in this case a spin-down torque due to gravitational waves, 
which is the main proposition of
\cite{HaskellPatruno2017}, would not be required anyway.
On the other hand, if pulsar magnetospheric activities are
quenched in the LMXB state, the $\gamma$-ray luminosity in this state
cannot be explained.
Therefore, we conclude that the gravitational wave emission {\it only} in the
LMXB state may not explain the torque budget and other properties
of J1023.
In fact, \cite{Singhetal2020} have recently shown that a large enough 
ellipticity, required to explain the increase of the spin-down rate
in the LMXB state of J1023, cannot be built only during the accretion phase,
unless strong shallow heating layers are present.

Recently, two models have been proposed to explain some properties of J1023, including 
the optical pulsations at the pulsar spin frequency 
\citep{Veledinaetal2019,Papittoetal2019}. Both these models assume that
the pulsar magnetosphere remains active in the LMXB state, the disc is truncated
outside the light cylinder in the high-mode, and optical and X-ray pulses 
are produced by synchrotron emission from the dissipative collision between a 
rotating pulsar wind and the inner edge cross section of the accretion disc. 
This mechanism is highly non-standard to explain the X-ray pulsations from 
an accreting MSP. Besides, an accretion disc outside the light cylinder 
radius ($\approx 80.5$~km) cannot explain a disc component found with 
$6.1\sigma$ significance, the inner edge radius of which has been inferred to be 
$21_{-7}^{+9}$~km (\cite{Campanaetal2016}; and $26_{-9}^{+11}$~km using later
data; \cite{CotiZelatietal2018}) from the high-mode X-ray spectrum (see 
section~\ref{Introduction}).

According to the model of \cite{Veledinaetal2019},
the accretion disc penetrates the light cylinder
during the low-mode in the LMXB state, but $r_{\rm m}$ remains greater than $r_{\rm co}$.
This can open up additional pulsar magnetic field lines and increase the spin-down torque
at most by a factor of $(r_{\rm lc}/r_{\rm m})^2$ \citep{Parfreyetal2016}.
This is the way \cite{Veledinaetal2019} attempt to explain the 
increase of the average spin-down rate in the LMXB state.
However for this, an evacuation of almost all the accreted matter from the light cylinder 
has to happen in $\sim 10$~s during the low--to--high mode transition,
so that the equatorial part of the pulsar wind is switched on to begin the high-mode,
according to this model.
Moreover, during every such transition, first the X-ray luminosity is expected to decrease
as the disc moves outwards, and then to 
sharply increase as the wind--disc collision happens.
On the other hand, during every high--to--low mode transition, the X-ray luminosity 
should first decrease sharply as the equatorial pulsar wind is shut down, 
and then somewhat increase as the disc extends inwards inside the magnetosphere. 
This could be tested using X-ray data with sufficient time resolution and statistics.

A main difference of the model of \cite{Papittoetal2019} with that of \cite{Veledinaetal2019} is
regarding the low-mode in the LMXB state.
According to \cite{Papittoetal2019}, the accretion disc is pushed away even farther
from the pulsar in the low-mode, without any change in the pulsar's magnetosphere and the accretion process.
Since, the pulsar magnetospheric activities should remain same in RMSP and LMXB states for this model, \cite{Papittoetal2019} would not perhaps
explain the significantly higher $|\dot{\nu}_{\rm LMXB}|$ value compared to the 
$|\dot{\nu}_{\rm RMSP}|$ value.

We, therefore, conclude that no existing model can explain the energy budget
and torque budget of J1023, using the standard scenario of X-ray pulsations 
in the LMXB state.

\section{Formalism and primary results}\label{Formalism}

In this section, we introduce our formalism using specific values of measured 
source parameters, as well as typical values of other
neutron star parameters.
This will be useful to demonstrate, in a simple manner, how this formalism works 
for J1023, and will probe if the neutron star emits gravitational waves.
In section~\ref{Method}, we will
describe the method to estimate the allowed ranges of several source
parameter values in a robust way, considering the full ranges of measured and
unknown input source parameter values.

\subsection{Implications of torque budget equations}\label{budget}

The energy budget and the torque budget of an MSP, which are related to energy and angular
momentum conservations respectively, are the most important observational physical aspects,
which any successful model must explain. 
Unlike some other models, our formalism for J1023 is based on this requirement.
Particularly, to the best of our knowledge, 
J1023 is so far the only MSP for which the time derivatives of the
spin frequency both in RMSP and LMXB states
have been measured. So, we start by utilizing this unique information, and
write the torque budget equations in the
two states. Note that, in the RMSP state, there could be only two types of torque:
due to pulsar wind and electromagnetic radiation from the pulsar magnetosphere
\citep{Lyne2012}
and gravitational waves from the pulsar \citep{Abbottetal2019}.
During the LMXB state, only one additional type of torque
can be present, i.e., due to the accreted/ejected matter. Since in the following
torque budget equations for J1023,
we include terms due to all these types of torque, these
are the most general torque budget equations without assumptions and irrespective
of other observational aspects.
Using the intrinsic values mentioned in section~\ref{Introduction}, 
we get for the RMSP state,
\begin{equation}\label{torque-RMSP}
        \dot{\nu}_{\rm RMSP}^{\rm PW} + \dot{\nu}_{\rm RMSP}^{\rm GW} = \dot{\nu}_{\rm RMSP} = -1.89\times10^{-15} ~{\rm Hz~s}^{-1},
\end{equation}

\noindent
and for the LMXB state,
\begin{equation}\label{torque-LMXB}
        \dot{\nu}_{\rm LMXB}^{\rm PW} + \dot{\nu}_{\rm LMXB}^{\rm GW} + \dot{\nu}_{\rm LMXB}^{\rm Acc} = \dot{\nu}_{\rm LMXB} = -2.50\times10^{-15} ~{\rm Hz~s}^{-1}.
\end{equation}

\noindent
Here, $\dot{\nu}$ is the rate of change of the neutron star spin frequency, which is related to the torque ($N$) by
$N = 2\pi\dot{\nu}I$, where $I$ is the stellar moment of inertia 
(section~\ref{Introduction}). The subscripts `RMSP' and `LMXB' indicate
the RMSP and LMXB states respectively, while the superscripts `PW', `GW' and `Acc' denote torques due to
pulsar magnetospheric activities, gravitational waves and accretion/ejection respectively.

In Eqs.~\ref{torque-RMSP} and \ref{torque-LMXB}, we include terms due to 
gravitational waves, i.e., $\dot{\nu}_{\rm RMSP}^{\rm GW}$ and 
$\dot{\nu}_{\rm LMXB}^{\rm GW}$. This is because a neutron star could support a certain 
amount of ellipticity around its spin axis \citep{JohnsonMcDaniel2013}, 
and hence could spin down due to gravitational wave emission. In fact, 
a population-based study of non-accreting MSPs indicated that such sources 
could have an ellipticity of $\sim 10^{-9}$ with respect to their 
spin-axis \citep{Woanetal2018}.
However, to the best of our knowledge, all the previous models for J1023 implicitly assumed
$\dot{\nu}_{\rm RMSP}^{\rm GW} = 0$, and almost all the previous models
\citep[except that of ][]{HaskellPatruno2017} implicitly assumed 
$\dot{\nu}_{\rm LMXB}^{\rm GW} = 0$, without providing a reason.
We make our formalism more complete by including these terms. 
But, we emphasize that we do not a priori assume non-zero values of
$\dot{\nu}_{\rm RMSP}^{\rm GW}$ and $\dot{\nu}_{\rm LMXB}^{\rm GW}$.
Rather, we estimate their values by solving Eqs.~\ref{torque-RMSP} and \ref{torque-LMXB}.
For this, let us first examine the other terms ($\dot{\nu}_{\rm LMXB}^{\rm Acc}$,
$\dot{\nu}_{\rm RMSP}^{\rm PW}$, $\dot{\nu}_{\rm LMXB}^{\rm PW}$)
in these equations.

The relatively low X-ray luminosity of J1023 implies a low accretion rate
and hence a low value of $|\dot{\nu}_{\rm LMXB}^{\rm Acc}|$ in Eq.~\ref{torque-LMXB}. While 
an unusual model of higher accretion rate, with $\sim 99$\% accreted matter ejected
from the system, was proposed mainly to explain the increased $\gamma$-ray emission
in the LMXB state \citep{PapittoTorres2015}, 
it has been argued in section~\ref{Previous} that an accretion rate much higher 
than the one indicated by the observed X-ray luminosity cannot generally explain the 
LMXB state of J1023. 
So, using a typical X-ray luminosity value of the LMXB state, 
$|\dot{\nu}_{\rm LMXB}^{\rm Acc}|$ can be
shown to be only a few percent of $|\dot{\nu}_{\rm LMXB}|$ in a simple way.
In sections~\ref{accretion} and \ref{Constraints}, 
we will estimate $\dot{\nu}_{\rm LMXB}^{\rm Acc}$
in more detail, averaging over all three modes, and find that it is positive and 
at most $\sim 5$\% of $|\dot{\nu}_{\rm LMXB}|$.

Next, we consider the terms $\dot{\nu}_{\rm RMSP}^{\rm PW}$ and
$\dot{\nu}_{\rm LMXB}^{\rm PW}$ in Eqs.~\ref{torque-RMSP} and \ref{torque-LMXB}.
As we will discuss in more detail in section~\ref{Feasibility},
$\gamma$-rays emitted in both RMSP and LMXB states of J1023 should be
powered by the pulsar spin energy through the pulsar wind and the electromagnetic 
radiation from the pulsar magnetosphere. This is the standard energy source of
a pulsar's $\gamma$-ray emission, and no other energy source for $\gamma$-rays is
known to be available in J1023. Note that, while the pulsar spin energy could
power the $\gamma$-rays in the LMXB state through $\sim 99$\% accreted matter ejected
from the system with high velocities \citep{PapittoTorres2015},
this model may not explain the LMXB state of J1023 (see section~\ref{Previous}).
Therefore, since the pulsar spin energy powers the $\gamma$-ray emission and
hence this emission partially causes the spin-down of the pulsar, 
$\dot{\nu}_{\rm RMSP}^{\rm PW}$ and $\dot{\nu}_{\rm LMXB}^{\rm PW}$ should be related 
to the $\gamma$-ray luminosities ($L_{\gamma, \rm{RMSP}}$ and $L_{\gamma, \rm{LMXB}}$) 
in RMSP and LMXB states of J1023, as shown by the next two equations.
The pulsar spin-down power in the RMSP state due to magnetospheric
outflow and emission is
\begin{equation}\label{LSDRMSP}
        L_{\rm SD, RMSP}^{\rm PW} = - 4 \pi^2 I \nu \dot{\nu}_{\rm RMSP}^{\rm PW} = L_{\gamma, \rm{RMSP}}/\eta_{\rm RMSP},
\end{equation}
where, $\eta_{\rm RMSP}$ is the $\gamma$-ray emission efficiency in the RMSP state.
Similarly, the pulsar spin-down power in the LMXB state
due to magnetospheric outflow and emission is
\begin{equation}\label{LSDLMXB}
        L_{\rm SD, LMXB}^{\rm PW} = - 4 \pi^2 I \nu \dot{\nu}_{\rm LMXB}^{\rm PW} = L_{\gamma, \rm{LMXB}}/\eta_{\rm LMXB},
\end{equation}
where, $\eta_{\rm LMXB}$ is the $\gamma$-ray emission efficiency in the LMXB state.
From Eqs.~\ref{LSDRMSP} and \ref{LSDLMXB}, we can write
\begin{equation}\label{keqn}
	\frac{L_{\rm SD, LMXB}^{\rm PW}}{L_{\rm SD, RMSP}^{\rm PW}}
	= \frac{\dot{\nu}_{\rm LMXB}^{\rm PW}}{\dot{\nu}_{\rm RMSP}^{\rm PW}}
= \frac{L_{\gamma, \rm{LMXB}}/L_{\gamma, \rm{RMSP}}}{\eta_{\rm LMXB}/\eta_{\rm RMSP}}.
\end{equation}

With this information about the terms $\dot{\nu}_{\rm LMXB}^{\rm Acc}$,
$\dot{\nu}_{\rm RMSP}^{\rm PW}$ and $\dot{\nu}_{\rm LMXB}^{\rm PW}$,
let us first examine from Eqs.~\ref{torque-RMSP} and \ref{torque-LMXB}
in an almost model-independent way, if $\dot{\nu}_{\rm RMSP}^{\rm GW}$ 
and/or $\dot{\nu}_{\rm LMXB}^{\rm GW}$ can be zero. This is 
because, if we can show in a model-independent way that at least 
one of them cannot be zero, then we can reliably infer a non-zero
ellipticity of and the gravitational wave emission from the neutron star.
We call this an almost model-independent way, 
because we use primarily some observational knowledge and 
the general torque budget equations (Eqs.~\ref{torque-RMSP} and 
\ref{torque-LMXB}), and these equations should be valid for any model.
Considering $L_{\gamma, \rm{LMXB}}/L_{\gamma, \rm{RMSP}} \ge 5$ (see
section~\ref{Introduction}) in Eq.~\ref{keqn}, 
it can be easily seen from Eqs.~\ref{torque-RMSP} and \ref{torque-LMXB}
that $|\dot{\nu}_{\rm RMSP}^{\rm GW}|$ has to be greater than zero
for $\dot{\nu}_{\rm LMXB}^{\rm GW} = 0$ and $\eta_{\rm LMXB}/\eta_{\rm RMSP} \lsim 3.5$, 
implying gravitational wave emission from the pulsar in the RMSP state.
Note that this upper limit of $\eta_{\rm LMXB}/\eta_{\rm RMSP}$,
for which $|\dot{\nu}_{\rm RMSP}^{\rm GW}|$ has to be greater than zero, is higher 
for $|\dot{\nu}_{\rm LMXB}^{\rm GW}| > 0$.

Since the conclusion of $|\dot{\nu}_{\rm RMSP}^{\rm GW}| > 0$ is valid only for a range of 
$\eta_{\rm LMXB}/\eta_{\rm RMSP}$, how reliable is it? In other words, 
what values of $\eta_{\rm LMXB}/\eta_{\rm RMSP}$ do we expect?
Note that, while the $\gamma$-ray emission should be 
powered by the pulsar spin energy through 
the pulsar wind and the electromagnetic radiation from the pulsar magnetosphere,
there could be multiple $\gamma$-ray production mechanisms. This means,
$\gamma$-rays may be produced either directly from the magnetospheric
outflow and emission (i.e., the $\gamma$-ray emission is a part of the pulsar
magnetospheric emission), or by an additional mechanism
(e.g., by a pulsar wind -- accretion disc collision; see section~\ref{Previous}), or both.
If the $\gamma$-ray emission in each of RMSP and LMXB states is
a part of the pulsar magnetospheric emission, and since
$L_{\gamma, \rm{LMXB}} > L_{\gamma, \rm{RMSP}}$, it is reasonable to
consider that $L_{\rm SD, LMXB}^{\rm PW} > L_{\rm SD, RMSP}^{\rm PW}$ for J1023
(section~\ref{gamma} will discuss a mechanism for 
$L_{\rm SD, LMXB}^{\rm PW} > L_{\rm SD, RMSP}^{\rm PW}$).
Hence, $\eta_{\rm LMXB}/\eta_{\rm RMSP} < 1$,
as the $\gamma$-ray efficiency is expected to decrease with the increase of the
pulsar spin-down power \citep{Lyne2012}.
Therefore, if the $\gamma$-ray emission in each of RMSP and LMXB states is
a part of the pulsar magnetospheric emission, then from the above it appears inevitable
that $|\dot{\nu}_{\rm RMSP}^{\rm GW}| > 0$, and the neutron star in J1023
has a non-zero ellipticity and emits gravitational waves in the RMSP state.

The other option, i.e., $\eta_{\rm LMXB}/\eta_{\rm RMSP} \ge 1$, 
typically implies that at least a fraction of $\gamma$-ray
emission in the LMXB state is produced by an additional mechanism (e.g., by a 
pulsar wind -- accretion disc collision; see section~\ref{Previous}). Therefore,
since $|\dot{\nu}_{\rm RMSP}^{\rm GW}| > 0$ for $\eta_{\rm LMXB}/\eta_{\rm RMSP} 
\lsim 3.5$, the neutron star in J1023 may emit gravitational waves in the RMSP state,
even if a considerable fraction of $\gamma$-rays in the LMXB state 
is produced by such an additional mechanism. 
Nevertheless, it is important to enquire if 
$\gamma$-rays in each of RMSP and LMXB states are actually
a part of the pulsar magnetospheric emission, because in this case 
the gravitational wave emission in the RMSP state appears to be certain.
While there is no independent observational evidence of such a direct pulsar magnetospheric
production of $\gamma$-rays in both states, in section~\ref{main}
we will argue that this production is the simplest explanation for $\gamma$-rays, has
not been falsified yet, is feasible, explains some properties
of J1023 better than alternative mechanisms, and also allows a self-consistent
scenario to understand the main observational aspects of J1023.

Note that, if the neutron star in J1023 has a
non-zero ellipticity in the RMSP state, it is unlikely to disappear in the LMXB state
(section~\ref{Constraints} includes a brief discussion). Therefore, if
$|\dot{\nu}_{\rm RMSP}^{\rm GW}| > 0$, $|\dot{\nu}_{\rm LMXB}^{\rm GW}|$ is
also expected to be greater than zero. 
In fact, considering ranges of parameter values and additional observational constraints,
we will report in section~\ref{Constraints} that both $|\dot{\nu}_{\rm RMSP}^{\rm GW}|$
and $|\dot{\nu}_{\rm LMXB}^{\rm GW}|$ are greater than zero.

\subsection{Results for specific parameter values}\label{Results}

Considering the above mentioned formalism and for typical input parameter values,
now we solve Eqs.~\ref{torque-RMSP}, \ref{torque-LMXB} and \ref{keqn},
in order to find out reasonable values of some source parameters.
We consider that $L_{\gamma, \rm{LMXB}}/L_{\gamma, \rm{RMSP}} = 6.5$
(see section~\ref{Introduction})
and $\eta_{\rm LMXB}/\eta_{\rm RMSP} = 1$.
We will explore reasonable ranges of these parameters 
in section~\ref{Method}. Therefore, from Eq.~\ref{keqn}, 
we get $\dot{\nu}_{\rm LMXB}^{\rm PW}/\dot{\nu}_{\rm RMSP}^{\rm PW} = 6.5$.
Here, we ignore the relatively small term $\dot{\nu}_{\rm LMXB}^{\rm Acc}$ 
in Eq.~\ref{torque-LMXB}, as this term does not affect the results much.
However, we will include this term in a self-consistent way in 
sections~\ref{Method} and \ref{Constraints}.
Here, we also consider $\dot{\nu}_{\rm LMXB}^{\rm GW} = \dot{\nu}_{\rm RMSP}^{\rm GW}$.
In sections~\ref{Method} and \ref{Constraints}, we will explore
the possibility of $\dot{\nu}_{\rm LMXB}^{\rm GW} \neq \dot{\nu}_{\rm RMSP}^{\rm GW}$.

Therefore, solving Eqs.~\ref{torque-RMSP} and \ref{torque-LMXB}, we get
$\dot{\nu}_{\rm RMSP}^{\rm PW} = -0.111\times10^{-15}$ Hz s$^{-1}$ and
$\dot{\nu}_{\rm RMSP}^{\rm GW} = \dot{\nu}_{\rm LMXB}^{\rm GW} = -1.779\times10^{-15}$ Hz s$^{-1}$.
The torque corresponding to $\dot{\nu}_{\rm RMSP}^{\rm PW}$ is given by \citep[see ][]{Lyne2012}
\begin{equation}\label{EMtorque}
        2\pi \dot{\nu}_{\rm RMSP}^{\rm PW} I = - \frac{2\mu^2\sin^2{\alpha}(2\pi\nu)^n}{3c^3},
\end{equation}
where $n$ and $\alpha$ are the pulsar braking index and the angle between the magnetic and spin axes, respectively.
The torque corresponding to $\dot{\nu}_{\rm RMSP}^{\rm GW}$ is given by \citep[see ][]{Bhattacharyya2017}
\begin{equation}\label{GWTorque}
        2\pi \dot{\nu}_{\rm RMSP}^{\rm GW} I = - \frac{32GQ^2}{5}\left(\frac{2\pi\nu}{c}\right)^5,
\end{equation}
where $Q$ is the misaligned mass quadrupole moment of the pulsar. 
Assuming typical parameter values, viz., 
$I = 2\times10^{45}$~g cm$^2$, $n = 3$ and $\alpha = 60^\circ$, we get $\mu \approx 3.8\times10^{25}$~G cm$^3$ and
$Q \approx 1.33\times10^{36}$~g cm$^2$ from Eqs.~\ref{EMtorque} and \ref{GWTorque}.
For a neutron star radius ($R$) of 14 km and the same $I$-value, these imply
the stellar surface dipole magnetic field $B = \mu/R^3 \approx 1.39\times10^7$~G and 
ellipticity $\epsilon = Q/I \approx 0.67\times10^{-9}$.

\section{Feasibility and implications}\label{Feasibility}

In this section, we briefly discuss the feasibility of our formalism 
(section~\ref{Formalism}), and its implications for various observational aspects.

\subsection{General aspects}\label{general}

First, we note that J1023 can be a spin-powered pulsar, if $B_{12}\nu^2 > 0.2$ 
\citep[$B_{12} = B/10^{12}$~G; ][]{BhattacharyaHeuvel1991}. 
This condition is well satisfied for the estimated $B$ value mentioned 
in section~\ref{Results}.
The estimated $Q$ and $\epsilon$ values are also consistent with the limits 
indicated for MSPs in \cite{Bhattacharyya2017, Woanetal2018}.

From the numbers mentioned in section~\ref{Results}, 
$L_{\rm SD, RMSP}^{\rm PW} = 5.2\times10^{33}$~erg s$^{-1}$
(using Eq.~\ref{LSDRMSP}). This may explain the observed irradiation of the companion star, which
requires a pulsar isotropic luminosity of at least $2.7\times10^{33}$~erg s$^{-1}$
(see section~\ref{Introduction}). The $\gamma$-ray emission
efficiency ($\eta_{\rm RMSP} = L_{\gamma, {\rm RMSP}}/L_{\rm SD, RMSP}^{\rm PW}$) in the RMSP state
is $\sim 0.23$ (from section~\ref{Results}), which is consistent with the 
values for the {\it Fermi}-LAT-detected MSPs \citep{Abdoetal2013}.

\subsection{Aspects of accretion}\label{accretion}

Let us now examine what our formalism and the corresponding reasonable parameter
values (see section~\ref{Results}) imply for the LMXB state of J1023.
Using $L_{\rm X} = 7\times10^{33}$~erg s$^{-1}$ (see section~\ref{Introduction}) 
and the usual factor of accreted mass to X-ray energy conversion (i.e., $GM/R$;
$M = 1.71 M_\odot$ and $R = 14$~km; sections~\ref{Introduction}
and \ref{Results}) in the high-mode, we get the accretion rate
$\dot{M} \sim 4.3\times10^{13}$~g s$^{-1}$. Then, using Eq.~\ref{rm}, $\xi = 0.3$
and $\mu = 3.8\times10^{25}$~G cm$^3$ (see section~\ref{Results}),
we get the magnetospheric radius $r_{\rm m, high}$
in the high-mode to be $\approx 24.6$~km. 
This is less than $r_{\rm co}$ ($=25.4$~km in this case;
Eq.~\ref{rco}), greater than both $R$ and $r_{\rm ISCO}$ ($\sim 15.4$~km), and hence
satisfies the condition for standard X-ray pulsation mechanism 
(see section~\ref{Previous}). Therefore, our formalism can explain
the X-ray pulsations of J1023 during the high-mode using the standard and 
well-accepted scenario, although this formalism is based on the torque
budget and $\gamma$-ray luminosities, and is independent of X-ray pulsations.
In fact, our formalism in section~\ref{budget} does not include the accretion
at all, except, based on a low observed X-ray luminosity, the  magnitude of the
accretion/ejection torque in the LMXB state is considered to be small in 
comparison to the total torque magnitude.
Besides, since the above values imply that the accreted matter falls on the 
neutron star surface in the high-mode, it is justified to use $GM/R$ as the
factor of accreted mass to X-ray energy conversion. It is also important to note that
the above $r_{\rm m, high}$ value is fully consistent with the disc inner edge radius
($r_{\rm in} = 21_{-7}^{+9}$~km; \cite{Campanaetal2016}; 
$r_{\rm in} = 26_{-9}^{+11}$~km; \cite{CotiZelatietal2018}) in the high-mode,
which was independently measured from X-ray spectral analysis. 
Therefore, the fact that our formalism can predict X-ray pulsations and 
$r_{\rm in}$ values, which are independently observed and measured
for the high-mode, strongly supports
this formalism and the conclusions described in section~\ref{Formalism}.

Here, we note that, unlike many other models, 
our formalism leads to the accretion onto the neutron star in the
high-mode, because the $\mu$-values in our scenario come out to be lower than those 
estimated assuming $\dot{\nu}_{\rm RMSP}^{\rm PW} = \dot{\nu}_{\rm RMSP}$. 
This is because we obtain a substantial non-zero value for 
$\dot{\nu}_{\rm RMSP}^{\rm GW}$, and hence
$|\dot{\nu}_{\rm RMSP}^{\rm PW}| < |\dot{\nu}_{\rm RMSP}|$ (see Eq.~\ref{torque-RMSP}) 
for our formalism.

Let us now consider the low-mode.
Using Eq.~\ref{rm}, the above mentioned parameter values (i.e., $M = 1.71 M_\odot$,
$\xi = 0.3$, $\mu = 3.8\times10^{25}$~G cm$^3$; sections~\ref{Introduction} 
and \ref{Results}), $L_{\rm X} = 10^{33}$~erg s$^{-1}$ 
(see section~\ref{Introduction}) and the factor of accreted mass
to X-ray energy conversion as $GM/2r_{\rm m, low}$ ($r_{\rm m, low}$ is the
magnetospheric radius in the low-mode), we find
$\dot{M} \sim 2.5\times10^{13}$~g s$^{-1}$ and $r_{\rm m, low} \approx 28.7$~km. 
This implies $r_{\rm lc} > r_{\rm m, low} > r_{\rm co}$ and J1023 is in 
the propeller regime in the low-mode (see section~\ref{Previous}), which 
explains why X-ray pulsations are not observed in this mode.
Moreover, since in this propeller regime the accreted matter should
not reach the neutron star surface, an energy conversion factor of
$GM/2r_{\rm m, low}$ is justified. 
Note that, had we used an energy conversion factor of $GM/R$,
$r_{\rm m, low}$ would be greater than $28.7$~km, and the accreted matter 
would not reach the stellar surface anyway. Therefore, $GM/R$ is not justified as
the energy conversion factor in the low-mode, as $GM/R$ can be used only 
when the matter falls on the stellar surface.
From the above we find that, although the solutions of Eqs.~\ref{torque-RMSP}, 
\ref{torque-LMXB} and \ref{keqn} do not use the accretion phenomenon apart from
considering the accretion/ejection torque magnitude to be small relative to 
the total torque magnitude, they lead to the conclusion that J1023 is in the
accretion regime in the high-mode and in the propeller regime in the low-mode. These two
different regimes naturally explain the two distinctly different modes 
(high and low) of the LMXB state of J1023.

As can be seen from the values of $r_{\rm m, high}$ and $r_{\rm m, low}$, a mode change
requires the disc to move only by a few km and to cross the $r_{\rm co}$. This can happen
in the viscous time scale $t_{\rm visc} \sim \alpha^{-1} (H/r)^{-2} (r^3/GM)^{1/2}$ \citep{Veledinaetal2019}.
Here, $\alpha$ is the Shakura-Sunyaev viscosity parameter and $H$ is the height of the disc at a radial distance $r$.
We consider $r \sim r_{\rm co} \sim 25$~km and $M \sim 1.71 M_\odot$, and find that
$t_{\rm visc} \sim 3 (\alpha/0.01)^{-1} (r/10H)^2$~s, which can comfortably explain the observed fast switch between modes (section~\ref{Introduction}).

The disc is truncated close to the $r_{\rm co}$, which is expected for a
mechanism discussed in \cite{DAngeloetal2010} 
\citep[also see ][]{BhattacharyyaChakrabarty2017}, and could explain the stability of
X-ray intensity in each of high- and low-modes. Since $r_{\rm co}$, being a function of
only $M$ and $\nu$ (Eq.~\ref{rco}), does not appreciably change within a few years,
the X-ray intensity in each mode could remain stable over the years, 
as observed \citep{Jaodandetal2016}.

The approximate magnitude of the torque ($|\dot{\nu}_{\rm LMXB}^{\rm Acc}|$; 
see Eq.~\ref{torque-LMXB}) due to accretion/ejection can be estimated from the formula
$|\dot M \sqrt{GMr_{\rm m}}|$ for each mode of the LMXB state \citep{BhattacharyyaChakrabarty2017}.
We consider a spin-up torque for the high and flaring modes,
with percentage durations of $\sim 77$\% and $\sim 1$\% respectively. For the low-mode,
we consider a spin-down torque using the above formula, and a percentage duration of $\sim 22$\%.
These give a net time derivative of spin frequency ($\dot{\nu}_{\rm LMXB}^{\rm Acc}$), 
averaged over the entire LMXB state, of $\sim 5.3\times10^{-17}$~Hz s$^{-1}$
(for parameter values mentioned in sections~\ref{Introduction} and \ref{Results}). 
This value of $\dot{\nu}_{\rm LMXB}^{\rm Acc}$ is only $\sim 2.1$\% of
$|\dot{\nu}_{\rm LMXB}|$, and hence its omission in a simple calculation is justified
(see section~\ref{Formalism}).
Note that, depending on the extent of the interaction of the neutron star 
magnetic field with the disc, and what fraction of the accreted matter can 
escape from the system by the propeller effect in the low-mode,
$\dot{\nu}_{\rm LMXB}^{\rm Acc}$ could have an uncertainty by a large fraction, 
but it would still remain a small fraction of $|\dot{\nu}_{\rm LMXB}|$.

\subsection{Aspects of $\gamma$-ray emission}\label{gamma}

Let us now discuss the feasibility of $\gamma$-ray emission being powered
by the pulsar spin energy through the pulsar magnetospheric activities in both
RMSP and LMXB states of J1023, as considered in our formalism (section~\ref{budget}).
In the RMSP state, the pulsar is clearly active, and hence $\gamma$-rays
can easily be powered by the pulsar spin energy through the magnetospheric activities.
In fact, this is the standard scenario of $\gamma$-ray origin for a radio MSP.
However, it is usually believed that pulsar magnetospheric activities are 
switched off due to accretion in an LMXB.
But, it has been seen from recent simulations that such activities 
could continue even when the accreted matter reaches the neutron star surface 
\citep{Parfreyetal2016,ParfreyetalTchekhovskoy2017}.
This can happen when a low accretion rate, corresponding to a low X-ray luminosity, cannot
quench the pulsar activities. We note that this could happen for J1023, 
as its average X-ray luminosity
in the LMXB state is less than $10^{34}$~erg s$^{-1}$, while the quenching luminosity was
estimated to be $\gsim 5\times10^{35}$~erg s$^{-1}$ \citep{CotiZelatietal2014}. 
Note that, for other AMXPs, the X-ray luminosity is typically higher than 
$5\times10^{35}$~erg s$^{-1}$ in the accretion phase, and hence the pulsar 
activities can be quenched. Therefore, the pulsar magnetospheric activities
can continue in the LMXB state of J1023, and the $\gamma$-ray emission powered
by the pulsar spin energy in this state is feasible.

But is it feasible that $\gamma$-rays are produced directly from 
the magnetospheric outflow and emission, i.e., the $\gamma$-ray emission is 
a part of the pulsar magnetospheric emission, in both RMSP and LMXB states of J1023?
For the RMSP state, this is feasible, and the $\gamma$-ray emission, 
similar to such emissions from many other MSPs detected with the {\it Fermi} Large 
Area Telescope (LAT), is in fact believed to be a part of the pulsar magnetospheric emission
\citep{Stappersetal2014}. But, if for the LMXB state of J1023 the $\gamma$-ray emission is
a part of this magnetospheric emission, then the observed $L_{\gamma, \rm{LMXB}} > 
L_{\gamma, \rm{RMSP}}$ indicates that the pulsar spin-down power 
($L_{\rm SD, LMXB}^{\rm PW}$) in the LMXB state is greater than that
($L_{\rm SD, RMSP}^{\rm PW}$) in the RMSP state (see section~\ref{budget}).
This can happen if there are more open magnetic field lines 
in the LMXB state than those in the RMSP state. But what can open more field lines
in the LMXB state?
As mentioned in section~\ref{Previous}, according to simulations, 
when the accretion disc extends inside the pulsar magnetosphere, at least a fraction of
the magnetic field lines intersecting the disc should be opened and remain opened 
\citep{Parfreyetal2016, ParfreyetalTchekhovskoy2017,Jaodandetal2016}. 
This should enhance the pulsar spin-down torque and spin-down power at most by a 
factor of $(r_{\rm lc}/r_{\rm m})^2$.
Note that, for $L_{\rm SD, LMXB}^{\rm PW} > L_{\rm SD, RMSP}^{\rm PW}$,
$\eta_{\rm LMXB}/\eta_{\rm RMSP} < 1$ (see section~\ref{budget}). Therefore,
as $L_{\gamma, \rm{LMXB}}/L_{\gamma, \rm{RMSP}}$ can be as high as $\sim 10$ for J1023
(although \cite{Delleretal2015} reported a lower value ($\sim 6.5$) of this ratio; 
section~\ref{Introduction}),
$L_{\rm SD, LMXB}^{\rm PW}$ may be sometimes more than 10 times 
$L_{\rm SD, RMSP}^{\rm PW}$ for this source (using Eq.~\ref{keqn}).
Can our formalism explain these for J1023? 
As shown in section~\ref{accretion}, our formalism
leads to the conclusion that the accretion disc extends inside the pulsar magnetosphere
in the LMXB state, and hence should enhance the spin-down power
at most by a factor of $(r_{\rm lc}/r_{\rm m})^2$,
which is $\approx 10.7$ (considering $r_{\rm m} \sim r_{\rm m, high} \approx 
24.6$~km) and $\approx 10$ (considering $r_{\rm m} \sim r_{\rm co} \approx
25.4$~km) using the example in section~\ref{accretion}.
Therefore, this formalism can explain the required increase of 
the pulsar spin-down power in the LMXB state.
So it is feasible for the $\gamma$-ray emission to be a part of the pulsar 
magnetospheric emission in the LMXB state of J1023.

\subsection{Main points of the formalism}\label{main}

We emphasize that our formalism and understanding of J1023 are based on three general
equations (\ref{torque-RMSP}, \ref{torque-LMXB} and \ref{keqn}).
Our main conclusion is, the gravitational wave emission from the pulsar
is inevitable for $\eta_{\rm LMXB}/\eta_{\rm RMSP} < k$, where
$k$ is somewhat greater than 1 (see section~\ref{budget}).
Therefore, as discussed before, if $\gamma$-rays are primarily
a part of the pulsar magnetospheric emission in both RMSP and LMXB states
(implying $\eta_{\rm LMXB}/\eta_{\rm RMSP} < 1$; section~\ref{budget}),
our conclusion of gravitational wave emission from the pulsar in J1023 should be
robust. While such a $\gamma$-ray production directly from the magnetospheric 
outflow and emission is natural and feasible in the RMSP state (see section~\ref{gamma}),
here we mention a few points to show that this production mechanism is quite likely 
even for the LMXB state.\\
(1) If the active magnetosphere of a pulsar can directly produce $\gamma$-rays in the 
RMSP state, then if this magnetosphere remains active in the LMXB state,
it might naturally keep on producing $\gamma$-rays in the latter state.
Then this would be the simplest mechanism for the $\gamma$-ray
production in the LMXB state, and an additional mechanism 
(e.g., by a pulsar wind -- accretion disc collision; see section~\ref{Previous})
would imply an additional model.\\
(2) Currently, no observation significantly falsifies the magnetospheric
production of $\gamma$-rays in the LMXB state. 
For example, no significant $\gamma$-ray spectral difference has been found 
between LMXB and RMSP states \citep{Stappersetal2014}.\\
(3) In section~\ref{gamma}, we have shown in detail that the magnetospheric 
$\gamma$-ray production mechanism in the LMXB state is feasible.\\
(4) While this production mechanism in our formalism can explain the energy 
budget, the torque budget and X-ray pulsations by the standard model, an 
alternative mechanism involving a pulsar wind -- accretion disc collision
may not explain all these (see section~\ref{Previous}).\\
Furthermore, our formalism does not require an additional $\gamma$-ray 
production mechanism (e.g., by a pulsar wind -- accretion disc collision),
and it self-consistently explains some main observational aspects of J1023
for a $\gamma$-ray production directly from the pulsar
magnetospheric outflow and emission. Here are two examples.\\
(1) Our formalism, based on Eqs.~\ref{torque-RMSP}, \ref{torque-LMXB} and \ref{keqn}
and a magnetospheric $\gamma$-ray production, infers accretion disc
penetration into the pulsar magnetosphere in the LMXB state and 
accretion onto the neutron star in the high-mode.
This, considering the opening of magnetic field lines, can naturally explain
an increased magnetospheric $\gamma$-ray production in the LMXB state.\\
(2) The equatorial part of the pulsar wind may be switched off by the
accretion disc penetrating the pulsar magnetosphere
\citep[see ][]{Veledinaetal2019} in the LMXB state.
Therefore, no significant collision between the pulsar wind and the accretion disc,
and hence no $\gamma$-ray production from such a collision, are expected in
our formalism. This is consistent with a magnetospheric $\gamma$-ray production.

\subsection{Other observational aspects}\label{other}

While it is not our aim to probe all the observational aspects of J1023, several such
aspects, as discussed above, can be comfortably explained using our formalism.
However, there are some observational aspects, which have not been reliably understood so far by any model,
and for these we can offer only speculative explanations.
Here are some examples.
The absence of observed radio pulsations in the LMXB state, when the pulsar magnetospheric activities
are present, could be due to intra-binary material causing scattering and/or absorption \citep{Stappersetal2014}.
The flat radio spectrum in the LMXB state indicated a jet \citep{Delleretal2015}, which, as
shown by simulations, could be due to the pulsar wind collimated by the accretion flow \citep{ParfreyetalTchekhovskoy2017}.
Besides, the optical pulsations at the pulsar spin frequency \citep{Ambrosinoetal2017}
could be explained in terms of the cyclotron emission 
from inflowing electrons in accretion columns within the accretion disc inner edge
\citep{Papittoetal2019}. But, the accretion columns
are expected to be optically thick, which would imply an optical luminosity more than an order
of magnitude lower than the observed optical pulsed luminosity \citep[$\sim 10^{31}$~erg s$^{-1}$; ][]{Papittoetal2019}.
We speculate that parts of the accretion columns are irradiated 
and puffed up by the energy from pulsar magnetospheric activities
and become optically thin. This might explain the optical pulsed luminosity
in terms of the cyclotron emission of electrons in accretion columns. Note that the corresponding energy budget
can comfortably be explained, as the optical pulsed luminosity is orders of magnitude
lower than the X-ray luminosity ($\sim 7\times10^{33}$~erg s$^{-1}$) due to accretion,
as well as the $\gamma$-ray luminosity ($\sim (6-12)\times10^{33}$~erg s$^{-1}$)
powered by the pulsar spin-down through magnetospheric emission.
This would also explain how the optical pulses originate within a few kilometers from the
location of the X-ray pulse origin and why the optical pulses are observed only in high- and flaring-modes (see section~\ref{Introduction}),
when accretion onto the pulsar happens according to our scenario. Moreover, optical pulsations have not been reported for other AMXPs yet, 
and hence this feature is likely uncommon for AMXPs,
perhaps because pulsar magnetospheric activities are quenched for them in
the accreting state (see section~\ref{gamma}).

\section{Method to constrain parameter values}\label{Method}

In sections~\ref{Formalism} and \ref{Feasibility}, we have presented our formalism
and primary results in a simple manner for typical parameter values.
In this section, we describe the method to constrain source parameter values,
including neutron star magnetic dipole moment and misaligned mass quadrupole moment,
following the formalism mentioned in section~\ref{Formalism} and the discussion
given in section~\ref{Feasibility}.
In order to make our results robust, we use full ranges of
measured parameter values and entire expected ranges of other input parameter values,
and include additional uncertainties for parameters such as accretion/ejection induced
rate of change of spin frequency.
To the best of our knowledge, hitherto a study of J1023 considering
such large ranges of parameter values has not been done.

The input parameters are pulsar mass $M$,
pulsar radius $R$, an order of unity constant $\xi$ (Eq.~\ref{rm}), $A$ (where, the pulsar moment of inertia
$I = AMR^2$), the $\gamma$-ray luminosity $L_{\gamma, \rm{LMXB}}$ in the LMXB state, the pulsar
braking index $n$, the angle $\alpha$ between the magnetic and spin axes (Eq.~\ref{EMtorque}),
$f_{\rm Acc}$ (uncertainty factor of $\dot{\nu}_{\rm LMXB}^{\rm Acc}$; Eq.~\ref{torque-LMXB}),
and $\eta_{\rm LMXB}/\eta_{\rm RMSP}$.
The ranges of values explored for these parameters are
$M: 1.55-1.87~M_\odot$ \citep{Delleretal2012};
$R: 8-16$~km \citep[the largest plausible range; e.g., ][]{Bhattacharyyaetal2017};
$\xi: 0.3-1.0$ \citep{PatrunoWatts2012};
$A: 0.33-0.43$ \citep{Bhattacharyyaetal2017};
$L_{\gamma, \rm{LMXB}}: (6.0-12.4)\times10^{33}$~erg s$^{-1}$ \citep{Nolanetal2012, Stappersetal2014, Delleretal2015, Torresetal2017};
$n: 1.4-3.0$ \citep{Lyne2012};
$\alpha: 5^{\circ}-90^{\circ}$ ($\sim$ entire range);
$f_{\rm Acc}: 0.5-1.5$ (a large fractional uncertainty) and
$\eta_{\rm LMXB}/\eta_{\rm RMSP}:~\le 3$ (a reasonably large range, including the
range for primarily pulsar magnetospheric origin of $\gamma$-rays).
In order to make our results reliable, we consider a large number
($\sim 1.8\times10^9$) of combinations of input parameter values within these ranges, and for each combination,
estimate various source parameter values, including stellar magnetic dipole 
moment $\mu$ and misaligned mass quadrupole moment $Q$, following the formalism 
described in section~\ref{Formalism} and the discussion given in section~\ref{Feasibility}.
In order to constrain the source parameter values, we impose the following 
conditions using observational aspects of J1023 (see section~\ref{Feasibility} for details).
(1) $r_{\rm m,high} \le r_{\rm co}$ and $r_{\rm m,low} \ge r_{\rm co}$, 
which are required to explain high- and low-modes, and the X-ray pulsations 
only in the high-mode using the standard mechanism.
(2) In the RMSP state, the pulsar magnetospheric spin-down luminosity $L_{\rm SD, RMSP}^{\rm PW} \ge
3\times10^{33}$~erg s$^{-1}$, which is required to explain the observed irradiation of the companion star.
(3) $(r_{\rm lc}/r_{\rm co})^2 > L_{\rm SD, LMXB}^{\rm PW}/L_{\rm SD, RMSP}^{\rm PW}$
(considering $r_{\rm m} \sim r_{\rm co}$), which is required for
magnetic field lines opening to explain the increased pulsar magnetospheric production of
$\gamma$-rays in the LMXB state.
In addition, following the theoretical structure computations of MSPs \citep{Bhattacharyyaetal2017}, we consider
$10^{45} {\rm g~cm}^2 \le I \le 4\times10^{45} {\rm g~cm}^2$.
The resulting constrained ranges of parameter values are reported in 
section~\ref{Constraints}.

\section{Constraints on parameter values}\label{Constraints}

\begin{table*}
\centering
	\caption{Constraints on parameters of PSR J1023+0038 (see sections~\ref{Method} and \ref{Constraints}).}
\begin{tabular}{|r|l|c|c|c|}
\hline
 & & & & \\
        No. & Parameters\footnotemark & For $\dot{\nu}_{\rm LMXB}^{\rm GW} = \dot{\nu}_{\rm RMSP}^{\rm GW}$ & For $\dot{\nu}_{\rm LMXB}^{\rm GW} \ge \dot{\nu}_{\rm RMSP}^{\rm GW}$ & No mutual condition \\
& & (with $\approx 0.2$\% accuracy) & & between $\dot{\nu}_{\rm LMXB}^{\rm GW}$, $\dot{\nu}_{\rm RMSP}^{\rm GW}$ \\
\hline
        1 & $r_{\rm co}$~(km) & $24.6-26.2$ & $24.6-26.2$ & $24.6-26.2$  \\ \\
        2 & Max$(R, r_{\rm ISCO})$~(km) & $14.0-16.8$ & $14.0-16.8$ & $14.0-16.8$ \\ \\
        3 & $r_{\rm m, high}$~(km) & $20.2-26.2$ & $19.4-26.2$ & $19.4-26.2$ \\ \\
        4 & $r_{\rm m, low}$~(km) & $24.6-32.5$ & $24.6-32.5$ & $24.6-32.5$ \\ \\
        5 & $(r_{\rm lc}/r_{\rm co})^2$ & $9.5-10.7$ & $9.5-10.7$ & $9.5-10.7$ \\ \\
        6 & $(r_{\rm lc}/r_{\rm m, high})^2$ & $9.5-15.9$ & $9.5-17.2$ & $9.5-17.2$ \\ \\
        7 & $\mu~(10^{25}$~G cm$^3$) & $2.5-4.4$ & $2.5-4.4$ & $2.5-4.4$ \\ \\
        8 & $\dot{\nu}_{\rm RMSP}^{\rm PW}$~($10^{-15}$~Hz s$^{-1}$) & $-0.067$ to $-0.217$ & $-0.033$ to $-0.217$ & $-0.033$ to $-0.217$ \\ \\
        9 & $\dot{\nu}_{\rm LMXB}^{\rm PW}$~($10^{-15}$~Hz s$^{-1}$) & $-0.692$ to $-0.941$ & $-0.066$ to $-0.941$ & $-0.066$ to $-2.243$ \\ \\
        10 & $\dot{\nu}_{\rm RMSP}^{\rm GW}$~($10^{-15}$~Hz s$^{-1}$) & $-1.673$ to $-1.823$ & $-1.673$ to $-1.857$ & $-1.673$ to $-1.857$ \\ \\
        11 & $\dot{\nu}_{\rm LMXB}^{\rm GW}$~($10^{-15}$~Hz s$^{-1}$) & $-1.670$ to $-1.827$ & $-1.673$ to $-2.484$ & $-0.295$ to $-2.484$ \\ \\
        12 & $\dot{\nu}_{\rm LMXB}^{\rm Acc}$~($10^{-15}$~Hz s$^{-1}$) & $0.015$ to $0.114$ & $0.014$ to $0.114$ & $0.014$ to $0.114$ \\ \\
        13 & $L_{\rm SD, RMSP}^{\rm PW}$ ($10^{33}$~erg s$^{-1}$) & $3.00-9.36$ & $3.00-9.36$ & $3.00-9.36$ \\ \\
        14 & $\eta_{\rm RMSP}$ & $0.13-0.40$ & $0.13-0.40$ & $0.13-0.40$ \\ \\
        15 & $Q_{\rm RMSP}$~($10^{36}$~g cm$^2$) & $0.92-1.88$ & $0.92-1.90$ & $0.92-1.90$ \\ \\
        16 & $Q_{\rm LMXB}$~($10^{36}$~g cm$^2$) & $0.92-1.88$ & $0.92-2.20$ & $0.39-2.20$ \\
\hline
\end{tabular}
\begin{flushleft}
$^1$Max$(R, r_{\rm ISCO})$ is the larger of $R$ and $r_{\rm ISCO}$. Parameters are defined in the text.\\
\end{flushleft}
\label{table_param}
\end{table*}

\begin{figure}
\centering
\hspace{-1.0cm}
\includegraphics*[width=9cm,angle=0]{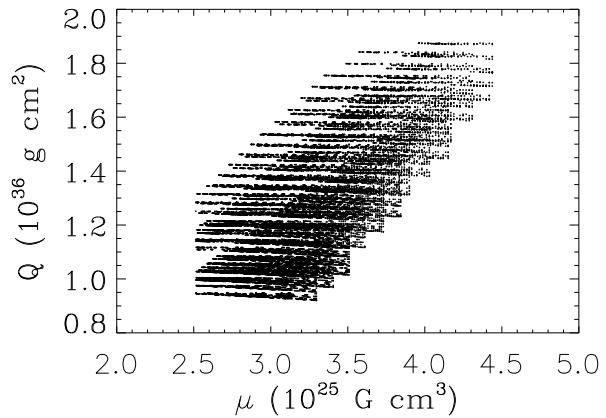}
        \caption{Constraint on the misaligned mass quadrupole moment ($Q$) versus
        magnetic dipole moment ($\mu$) space for the pulsar PSR J1023+0038
	(see section~\ref{Constraints}). Here, we use full ranges of
measured parameter values, entire expected ranges of other input parameter values,
	and a few additional constraints (see section~\ref{Method} for details).
        Besides, here we consider that $Q$ is same in RMSP and LMXB states.
        The clustering of points is due to the chosen grid of parameters.
        The overall increase of $Q$ with $\mu$ is due to the increase of the pulsar
        moment of inertia $I$, which is an input parameter (see Eqs.~8 and 9). For a given $I$-value, however,
        there is typically a weak anticorrelation between $Q$ and $\mu$, which is expected
        (see Eqs.~3, 8 and 9).
\label{fig1}}
\end{figure}

Using our formalism (section~\ref{Formalism}) and discussion in section~\ref{Feasibility},
full ranges of measured parameter values, and entire expected ranges of other input 
parameter values (see section~\ref{Method}), we constrain a number of parameters 
for J1023, and list their allowed values in Table~\ref{table_param}.
Such a systematic and extensive study of the parameter space of J1023 has not been done so far, to the best of our knowledge.
Theoretical calculations predict that a neutron star could sustain a $Q$ and an
$\epsilon$, which are orders of magnitude higher than $\sim 10^{36}$~g cm$^2$ and
$\sim 10^{-9}$ respectively \citep{JohnsonMcDaniel2013}.
Therefore, as the stellar gravity cannot iron out such an unevenness, structural 
imperfections should remain, and hence a net $\epsilon \sim 10^{-9}$ could 
naturally exist in neutron stars. 
For example, a buried strong magnetic field or a strain in the elastic crust could
give rise to a non-zero ellipticity \citep{Woanetal2018}.
Moreover, if such a natural ellipticity exists in the RMSP state of J1023, it should not
disappear in the LMXB state. Therefore, without assuming an additional source of ellipticity in the latter
state, we consider $\dot{\nu}_{\rm LMXB}^{\rm GW} = \dot{\nu}_{\rm RMSP}^{\rm GW}$ (implying
$Q_{\rm LMXB} = Q_{\rm RMSP} = Q$ and $\epsilon_{\rm LMXB} = \epsilon_{\rm RMSP} = \epsilon$)
for the first set of constraints in Table~\ref{table_param}.
We find $\mu \approx (2.5-4.4)\times10^{25}$~G cm$^3$, implying $B \approx (0.7-6.1)\times10^{7}$~G, and
$Q \approx (0.92-1.88)\times10^{36}$~g cm$^2$, implying $\epsilon \approx 
(0.48-0.93)\times10^{-9}$.
The constrained $Q-\mu$ space is shown in Fig.~\ref{fig1}.

While $Q_{\rm LMXB} = Q_{\rm RMSP}$ is the simplest possibility, 
let us now consider more general cases.
If the pulsar ellipticity can increase in the LMXB state due to an additional 
accretion-related asymmetry, which requires an additional mechanism
(see \cite{HaskellPatruno2017,Singhetal2020} for an example of a mechanism), 
then we get $Q_{\rm RMSP} \approx (0.92-1.90)\times10^{36}$~g cm$^2$ and
$Q_{\rm LMXB} \approx (0.92-2.20)\times10^{36}$~g cm$^2$ (Table~\ref{table_param}).
Therefore, our formalism does not exclude a model of the
pulsar ellipticity increase during accretion.
However, depending on the relative orientations of the natural 
ellipticity and the accretion-related asymmetry,
the net ellipticity could even decrease in the LMXB state.
Therefore, without considering any mutual condition between $\dot{\nu}_{\rm RMSP}^{\rm GW}$ and
$\dot{\nu}_{\rm LMXB}^{\rm GW}$, we find  $Q_{\rm RMSP} \approx (0.92-1.90)\times10^{36}$~g cm$^2$ and
$Q_{\rm LMXB} \approx (0.39-2.20)\times10^{36}$~g cm$^2$ (Table~\ref{table_param}).
These may imply that J1023 has a permanent ellipticity, and emits gravitational waves 
in both RMSP and LMXB states, even when we
do not impose any constraints on $\dot{\nu}_{\rm RMSP}^{\rm GW}$ and
$\dot{\nu}_{\rm LMXB}^{\rm GW}$.

What is the possibility of detection of gravitational waves from J1023?
For MSPs ($\nu > 100$~Hz) with ellipticities of $10^{-9}$ and the
canonical moment of inertia ($I$), and for a one-year observation period with 
coherently combined data, \cite{Woanetal2018} have estimated the signal-to-noise
ratios. We find that the 
ellipticity of the neutron star in J1023 could be somewhat less than 
$10^{-9}$, but its $\nu$-value is among the highest values observed for MSPs.
Moreover, considering the numerically and general relativistically 
computed $I$-values for wide varieties of equation of state models
for an MSP (PSR J1903+0327) of a comparable mass 
\citep[see Table 3 of ][]{Bhattacharyyaetal2017}, we may assume that
the $I$-value of J1023 could be around $(1.7-3)\times10^{45}$~g cm$^2$,
which is relatively high. Therefore, considering Fig.~2 of
\cite{Woanetal2018}, while it is unlikely that gravitational waves from J1023
would be detected with two upgraded advanced LIGO detectors,
together with the advanced Virgo detector \citep{Acerneseetal2015}, 
there could be a reasonable chance 
of detection with the Cosmic Explorer \citep{Abbottetal2017}
and the Einstein Telescope in its ET-D configuration 
\citep{Hildetal2011,Sathyaprakashetal2012}.
Note that a plausible detection of continuous gravitational waves from J1023
and the corresponding estimation of $Q$ in the future may confirm
our formalism, infer some source parameter values using our method,
and generally be useful to probe the physics of neutron stars, binary 
systems, accretion through a magnetosphere and stellar and binary evolutions.

\section{Summary}\label{Summary}

In this section, we summarize the main points of this paper.\\
(1) A millisecond pulsar (MSP) could sustain a misaligned mass distribution or ellipticity
around the spin-axis, and hence could emit continuous gravitational waves.
An indirect way to infer such waves is to estimate the contribution of
the waves to the spin-down rate of the pulsar.
The transitional pulsar PSR J1023+0038 is so far the only MSP for which this
spin-down rate has been measured in both the non-accreting radio MSP (RMSP) state
and the accreting low-mass X-ray binary (LMXB) state. This unique information
makes PSR J1023+0038 ideal to theoretically investigate if the pulsar has
a non-zero ellipticity.\\
(2) While we do not a priori assume a non-zero ellipticity of this pulsar,
we can show that torque budget equations and pulsar magnetospheric origin of
observed $\gamma$-rays in two states naturally infer such an ellipticity and 
continuous gravitational waves from PSR J1023+0038.\\
(3) Our formalism, which is basically independent of accretion, 
infers accretion disc penetration into the pulsar magnetosphere,
and explains the observed X-ray pulsations in the LMXB state using the standard 
and well-accepted scenario. This, in turn, naturally infers the significantly larger pulsar
spin-down power in the LMXB state, which we use in our formalism to explain 
the observed larger $\gamma$-ray emission in this state. This shows the 
self-consistency of our formalism.\\
(4) Finally, exploring large ranges of parameter values, and without assuming an
additional source of ellipticity in the LMXB state, we infer a permanent 
misaligned mass quadrupole moment of the pulsar in the range 
$(0.92-1.88)\times10^{36}$~g cm$^2$, implying the $(0.48-0.93)\times10^{-9}$
range of ellipticity for PSR J1023+0038.
 

\section{Data Availability}

Observational data used in this paper are publicly available, and references are mentioned. 
The data generated from computations are already reported in this paper, and any 
additional used data will be available on request.



\bibliography{ms}

\bibliographystyle{mnras}

\bsp    
\label{lastpage}
\end{document}